\documentclass[amsmath, amssymb,jcp, twocolumn, notitlepage, 10pt, superscriptaddress]{revtex4-1}
\usepackage{times}
\usepackage{float}
\usepackage{fancyhdr}
\fancyheadoffset[L,R]{\headrulewidth}
\renewcommand{\headrulewidth}{0.6pt}

\usepackage{cancel}
\usepackage[normalem]{ulem}

\usepackage{cancel}
\usepackage{mathrsfs}
\usepackage{graphicx}
\usepackage{tikz,pgf}
\usepackage{color}
\usepackage{array}
\usepackage{xcolor}
\usepackage[dvips]{epsfig}
\usepackage{epsfig}
\usepackage{enumerate}
\usepackage[latin1]{inputenc}
\usepackage[T1]{fontenc}
\usepackage[hang]{subfigure}
\usepackage{bbm, dsfont}
\usepackage{amsfonts,stmaryrd}
\usepackage{ulem}

\usepackage{bbold}

\newcommand{\bra}[1]{\langle #1 | \,}
\newcommand{\ket}[1]{\, | #1 \rangle}
\newcommand{\braket}[2]{\langle #1 | #2 \rangle}

\newcommand{\ga}{\ga}

\newcommand{\bl}{\begin{linenomath*}}
\newcommand{\el}{\end{linenomath*}}

\newcommand{\bea}{\begin{eqnarray}}
\newcommand{\eea}{\end{eqnarray}}
\renewcommand{\vec}[1]{\mathbf{#1}}

\renewcommand{\ga}{\hat\gamma}

\definecolor{dgreen}{rgb}{0.0, 0.5, 0.0}

\usepackage{hyperref}

\begin{document}

%%%%%%%%%%%%%%%%%%%%%%%%%%%%%%%%%%%%%%%%%%%%%%%%%%

 \title{Effect of a magnetic field on molecule--solvent angular momentum transfer}

%%%%%%%%%%%%%%%%%%%%%%%%%%%%%%%%%%%%%%%%%%%%%%%%%%
\author{Wojciech Rz\k{a}dkowski}
\email{wojciech.rzadkowski@ist.ac.at}
\affiliation{IST Austria (Institute of Science and Technology Austria), Am Campus 1, 3400 Klosterneuburg, Austria}
\affiliation{Faculty of Physics, University of Warsaw, ul. Pasteura 5, 
02-093 Warszawa, Poland}
\author{Mikhail Lemeshko}
\email{mikhail.lemeshko@ist.ac.at}
\affiliation{IST Austria (Institute of Science and Technology Austria), Am Campus 1, 3400 Klosterneuburg, Austria}

\begin{abstract}
Recently it was shown that a molecule rotating in a quantum solvent can be described in terms of the `angulon' quasiparticle [Phys.\ Rev.\ Lett.\ \textbf{118}, 095301 (2017)]. Here we extend the angulon theory to the case of molecules possessing an additional spin--$1/2$ degree of freedom and study the behavior of the system in the presence of a static magnetic field. We show that exchange of angular momentum between the molecule and the solvent can be altered by the field, even though the solvent itself is non-magnetic. In particular, we demonstrate a possibility to control resonant emission of phonons with a given angular momentum using a magnetic field.
\end{abstract}

\maketitle

\section{Introduction}

Quite often, properties of quantum many-particle systems can be understood by 
considering their elementary building blocks -- individual impurities coupled to 
a many-particle environment~\cite{Mahan90, WeissBook, Breuer2002}.  During the 
last years much  effort has been focused on uncovering the physics associated 
with point-like impurities, possessing simple or no internal structure. Such 
impurity problems  date back to the concept of polaron first introduced by 
Landau and Pekar~\cite{LandauPolaron, Pekar46, LandauPekarJETP48}. The polaron represents a quasiparticle consisting of an electron 
dressed by a cloud of crystal vibrations, and has become a standard  tool to 
describe transport phenomena in solid state and chemical 
physics~\cite{EminPolarons, Devreese15}. 
Recently, controllable polarons have been realized in ultracold quantum gases of 
bosons and fermions~\cite{MassignanRPP14, Jorgensen2016}. 
Another broad class of well-studied impurity problems involves a localized spin 
coupled to a bath of bosons~\cite{LeggettRMP87}, fermions~\cite{LutchynPRB08}, 
or other spins~\cite{ProkofievSpinBath00}.

In many settings, however, quantum impurities possess additional internal 
degrees of freedom, such as orbital or rotational angular momentum. Such problems arise, e.g.\ 
in the context of molecules rotating in superfluid 
helium~\cite{ToenniesAngChem04}, ultracold alkali dimers interacting with a 
Bose-Einstein condensate (BEC)~\cite{JinYeCRev12}, or electrons whose orbital 
angular momentum is coupled to the crystal lattice~\cite{KoopmansPRL05, 
ChudnovskyPRB05, Zhang2014, Tows2015, Garanin2015, Tsatsoulis2016}. Recently, 
one of us has demonstrated that problems of that sort can be conveniently 
addressed by introducing a new species into the quasiparticle 
zoo -- the  `angulon' quasiparticle~\cite{SchmidtLem15, 
SchmidtLem16, LemSchmidtChapter}. The angulon forms out of an impurity 
exchanging rotational angular momentum with a many-particle bath of some sort; it 
can be thought of as a quantum rotor dressed in a coat of orbital bath 
excitations. In a way, the angulon represents a rotational counterpart of the 
polaron, however, the non-abelian algebra of quantum rotations and their 
discrete spectrum render the angulon physics remarkably 
different.

The angulon theory has been tested against experiments on molecules trapped in 
superfluid helium nanodroplets. There, it was observed that the effective moment 
of inertia increases for molecules immersed in superfluid helium, as compared to 
free species~\cite{ToenniesAngChem04}. This phenomenon is somewhat similar to 
renormalization of the effective mass of electrons interacting with a crystal 
lattice~\cite{Devreese15}. It 
was recently shown that the angulon theory can reproduce the effective moments 
of inertia for molecules in helium nanodroplets for a broad range of species, 
both in the weak-coupling and strong-coupling regimes~\cite{LemeshkoDroplets16}. Moreover, a coherent non-adiabatic rotation of angulons formed 
out of I$_2$ molecules has been experimentally demonstrated~\cite{Shepperson16}. The angulon theory thereby offers an alternative approach to molecules in quantum solvents, along with established numerical techniques based on quantum Monte Carlo and Density Functional Theory calculations~\cite{SzalewiczIRPC08, RodriguesIRPC16, AncilottoIRPC17}.

The concept of angulons allowed to predict a number of novel physical phenomena 
associated with rotating impurities. As an example, it was demonstrated that for 
some values of impurity-bath coupling, orbital angular momentum can be 
resonantly transferred from the impurity to the bath, resulting in the `angulon 
instabilities'. The fingerprints of these instabilities 
were identified in infrared spectra of molecules in helium 
nanodroplets~\cite{Cherepanov, MorrisonJPCA13, Slipchenko2005}. Remarkably, such resonant transfer 
of angular momentum leads to anomalous screening of the impurities' dipole 
moments and polarizabilities, even if they reside in a neutral, non-polarizable 
environment~\cite{Yakaboylu16}. Furthermore, angulon spectra reveal the 
rotational Lamb shift  -- the phononic analogue of the Lamb shift in hydrogen, 
as described by quantum electrodynamics~\cite{Rentrop16, 
ScullyZubairy}.

%The rotational Lamb shift was shown to alter the spectra of light molecules in superfluid helium~\cite{LemeshkoDroplets16} and was predicted to be experimentally detectable for ultracold molecular ions interacting with a BEC~\cite{Midya2016}. Later, it was shown that in the presence of a strong electrostatic field, angulons turn into `pendulons' -- spherical harmonic librators dressed by many-particle excitations~\cite{Redchenko16}. Finally, the possibility to observe angular self-localization transition in angulons has also been studied~\cite{Li16}.

In this paper we generalize the angulon theory to the case where the impurity 
possesses both rotational and spin--$1/2$ degrees of freedom and is exposed to a 
static magnetic field. Our main focus will be on open-shell diatomic molecules rotating in quantum solvents. 
{It is important to emphasize that the angulon model (including Eqs.~\eqref{Himpbos} and \eqref{eq:dispersion} below) has been originally derived for an ultracold molecule immersed in a weakly-interacting BEC, where the theory is expected to provide quantitatively accurate predictions. It has been shown, however, that one can approach the angulon Hamiltonian from a phenomenological perspective in order to describe the properties of molecules in superfluid $^4$He, in good agreement with experiment~\cite{LemeshkoDroplets16, Cherepanov, Shepperson16}. Thus, while the theory is not designed to compete with numerical Monte Carlo calculations in accuracy, it is expected to provide qualitatively accurate predictions for molecules in liquid helium along with simple explanations for the underlying physics.} Furthermore, the theory can be in principle generalized to other types of orbital impurities such as polyatomic molecules, non-spherical paramagnetic atoms, or $p-$, 
$d-$, or $f-$electrons. The paper is organized as follows. In Sec.~\ref{sec:theory} we 
derive the extended angulon Hamiltonian, which includes the impurity spin and 
the impurity-field interaction. In Sec.~\ref{sec:variational} we introduce an 
approach to the extended Hamiltonian, based on variational and diagrammatic 
techniques. In Sec.~\ref{sec:zeeman} we analyze the angulon spectral function 
and the way it changes in a magnetic field, for various bath densities.  
Sec.~\ref{sec:manipulating} focuses on the angulon instabilities which result in 
resonant emission of phonons with a given value of angular momentum. In 
particular, we reveal the possibility to manipulate the angular momentum of 
phonons using a magnetic field. The conclusions and outlook of this paper are 
presented in Sec.~\ref{sec:discussion}.

\section{Spinful molecular angulons in a magnetic field}
\label{sec:theory}

We consider a molecular impurity with spin--$1/2$ and  orbital angular momentum, immersed in a bosonic bath. In the presence of a magnetic field, the system can be described by the following Hamiltonian:
\begin{equation}
\label{hamiltonian}
\widehat{H}=\widehat{H}_{\textnormal{mol}}+\widehat{H}_{\textnormal{mol-f}}
+\widehat { H }_{\textnormal{bos}}+ \widehat { H }_{\textnormal{mol-bos}},
\end{equation}
where the terms correspond to the bare molecule, molecule-field interaction, bosonic bath, and molecule-boson interaction, respectively. Note that we assume a neutral, spinless bath, such that its direct interaction with the magnetic field can be neglected. In the following subsections we describe each term of Eq.~\eqref{hamiltonian} in detail. In what follows, we use the units where $\hbar \equiv 1$.

\subsection{Bare spinful molecule}
The first term of Eq.~\eqref{hamiltonian} corresponds to a linear molecule with spin--$1/2$ ($^2\Sigma$ electronic state), as given by the following Hamiltonian:
\begin{equation}
\label{Himp}
 \widehat{H}_{\textnormal{mol}}= B \hat{\vec{L}}^2 + 
\gamma\hat{\vec{L}}\cdot\hat{\vec{S}}.
\end{equation}
Here $B = 1/(2I)$ is the rotational constant with $I$ being the 
moment of inertia, and $\gamma$ defines the spin-rotation 
coupling~\cite{LevebvreBrionField2}.

Since we focus on spin--$1/2$ molecules, in Eq.~\eqref{Himp} we omit the 
constant shift proportional to $\hat{\vec{S}}^2$. For higher spins (as,  e.g., in $^3\Sigma$ molecules), the 
spin-spin interaction will lead to  an additional term in $ 
\widehat{H}_{\textnormal{mol}}$. Furthermore, here we consider an impurity whose 
translational motion is frozen in space, which is a good approximation for molecules in helium 
nanodroplets~\cite{ToenniesAngChem04}. Our formalism, however, can be 
generalized to include the above mentioned terms, as well as to treat more 
complex impurities, such as polyatomic molecules~\cite{BernathBook}.
 
We denote the eigenstates of the bare molecular Hamiltonian~(\ref{Himp}) 
as $\ket{J=L\pm1/2,L,M_J}$ with the corresponding eigenenergies
\begin{equation}
\label{bareenergies}
E^0_{JL}= BL(L+1)+
\gamma\left[(J-L)\left(L+\frac{1}{2}\right)-\frac{1}{4}\right].
\end{equation}

Here $J$ is the total (rotation+spin) angular momentum, $L$ is the total 
rotational angular momentum and $M_J$ is the projection of $J$ on the 
quantization axis. All three numbers are good quantum numbers. Note that the 
eigenstates can be written in the uncoupled basis as
\begin{equation}
\label{barestates}
\begin{split}
&\ket{J=L\pm\frac{1}{2},L,M_J}=\\
&C^{J,M_J}_{L,M_J-\frac{1}{2};\frac{1}
{ 2 }
, \frac { 1 } { 2 } }\ket{L,M_J-1/2}\ket{\frac{1}{2},\frac{1}{2}}\\
&+
 C^{J,M_J}_{L,M_J
 +\frac{1}{2};\frac{1}
 { 2 }
 , -\frac{ 1 }{ 2 } }\ket{L,M_J+1/2}\ket{\frac{1}{2},-\frac{1}{2}}.
\end{split}
\end{equation}

\subsection{Molecule-field interaction}
The therm $\widehat{H}_{\textnormal{mol-f}}$ describes a static magnetic field 
applied to the system:
\begin{equation}
\label{eq:Himpf}
\widehat{H}_{\textnormal{mol-f}}= B\eta\hat{S}_Z,
\end{equation}
where the dimensionless molecule-field interaction parameter is given by:
\begin{equation}
\eta=\frac{\mathscr{H}g_s\mu_B}{B}. 
\end{equation}
Here $g_s\approx2.0023$ is the gyromagnetic ratio, $\mu_B$ is the
Bohr magneton, and $\mathscr{H}$ gives the magnitude of the magnetic field.

Assuming that the magnetic field affects only the spin degree of freedom 
(i.e.\ neglecting rotational magnetism in the case of 
molecules~\cite{LevebvreBrionField2}), the operator~\eqref{eq:Himpf} couples 
only levels with the same $L$.
 In such a case, the eigenstates of the molecule+field Hamiltonian,  
$\widehat{H}_{\textnormal{mol}}+\widehat{H}_{\textnormal{mol-f}}$, represent 
field-dependent superpositions of 
the states $ \ket{J,L,M_J}$~\cite{Friedrich2000}: 
\begin{equation}
\label{superposition}
\begin{split}
\ket{\tilde{J},L,M_J}= & a_{\tilde{J}LM_J}(\eta)\ket{J=L+1/2,L,M_J}+\\
+&b_{\tilde{J}LM_J}(\eta)\ket{J=L-1/2,L,M_J}.
\end{split}
\end{equation}
In the presence of a field, $L$ and $M_J$ are good quantum numbers, while $J$ is not. However, in Eq.~\eqref{superposition} we use $\tilde{J}$ as an adiabatic (approximately good) quantum number, such that

\begin{equation}
\ket{\tilde{J},L,M_J} \underset{\eta \rightarrow 0}{\longrightarrow} 
\ket{J,L,M_J}. 
\end{equation}
The exact form of the $a_{\tilde{J}LM_J}(\eta)$ and $b_{\tilde{J}LM_J}(\eta)$ 
coefficients of Eq.~\eqref{superposition} is given in 
Appendix~\ref{appendix_details}.

The eigenenergies of the $\ket{\tilde{J},L,M_J}$ states are given 
by~\cite{Friedrich2000}:
\begin{equation}
\label{eigenenergy}
E^0_{\tilde{J}LM_J}=
BL(L+1)-\frac{\gamma}{4}
+(\tilde{J}-L)\gamma(L+\frac{1}{2})
\xi_{BLM_J},
\end{equation}
where 
\begin{equation}
\label{eigenenergy1}
 \xi_{BLM_J}=(1+2\alpha_{LM_J}X_{BL}+X_{BL}^2)^{\frac{1}{2}},
\end{equation}
with
\begin{equation}
\label{eigenenergy2}
 \alpha_{LM_J}=\frac{M_J}{L+1/2},\hspace{0.1cm} X_{BL}=\frac{B\eta}{\gamma(L+1/2)}.
\end{equation}

\subsection{Bosonic bath energy}
The term $\widehat{H}_{\textnormal{bos}}$ corresponds to the kinetic energy of the bosonic excitations in a quantum solvent, such as phonons,   rotons, and ripplons in superfluid 
$^4$He~\cite{StienkemeierJPB06}. In its diagonal form, the bosonic bath 
Hamiltonian reads:
\begin{equation}
\label{Hbos}
\widehat{H}_{\textnormal{bos}}=\sum\limits_{k\lambda\mu} \omega_k 
\hat{b}^\dag_{k\lambda\mu} 
\hat{b}_{k\lambda\mu},
\end{equation}
where $\omega_k$ is the dispersion relation. The creation and annihilation operators of Eq.~\eqref{Hbos} are conveniently expressed in 
the angular momentum basis~\cite{LemSchmidtChapter}:
\begin{equation}
\hat{b}^\dagger_{k\lambda\mu}= \frac{k}{(2 \pi)^{3/2}} \int d\Omega_k~
i^{-\lambda}~
Y_{\lambda\mu}(\Omega_k)~\hat{b}^\dagger_{\mathbf{k}},
\end{equation}
\begin{equation}
\hat{b}_{k\lambda\mu}= \frac{k}{(2 \pi)^{3/2}} \int d\Omega_k~i^{\lambda}~Y^\ast_{\lambda\mu}(\Omega_k)~\hat{b}_{\mathbf{k}}.
\end{equation}
Here $\hat{b}^\dagger_{\mathbf{k}}$ and $\hat{b}_{\mathbf{k}}$ are the creation and annihilation operators defined in  Cartesian space, and $Y_{\lambda\mu}(\Omega_k)\equiv Y_{\lambda\mu}(\theta_k,\varphi_k)$ are the spherical harmonics~\cite{Varshalovich}. The quantum numbers $k=|\mathbf{k}|$, $\lambda$, and $\mu$, label, respectively, the linear momentum of phonons, the angular momentum of phonons, and the projection of the phonon angular momentum onto the laboratory-frame $z$-axis.

In Eq.~\eqref{Hbos}, the form of the dispersion relation $\omega_k$ depends on 
the particular system under consideration. Here, without loss of generality, we 
chose the dispersion relation corresponding to Bogoliubov excitations in a 
weakly-interacting BEC~\cite{Pitaevskii2016}:
\begin{equation}
\omega_k=\sqrt{\epsilon_k(\epsilon_k+2g_{bb}n)}.
\label{eq:dispersion}
\end{equation}
Here $\epsilon_k= k^2/(2m)$ is the boson kinetic energy and $g_{bb}=4\pi a_{bb}/m$ parametrizes the interactions between the bosons of mass $m$, where $a_{bb}$ gives the boson-boson scattering length. While the Bogoliubov dispersion~\eqref{eq:dispersion} does not provide a quantitatively good approximation to the properties of superfluid helium, in the regime of large values of $g_{bb}$ and $n$ it qualitatively describes the properties of a dense superfluid for small momenta $k$. Furthermore, the theory can be extended to other types of excitations, such as rotons or lattice phonons.

\subsection{Molecule-boson interaction}
The last term of Eq.~\eqref{hamiltonian} determines the   interaction between the  molecule and the bosonic bath, as given by~\cite{SchmidtLem15}:
\begin{equation}
\label{Himpbos}
 \widehat{H}_{\textnormal{mol-bos}}=\sum\limits_{k\lambda\mu} U_\lambda(k) ~ 
 [Y^*_{\lambda\mu}(\hat{\theta},\hat{\phi})\hat{b}^\dag_{k\lambda\mu}+Y_{
 \lambda\mu}(\hat{\theta},\hat{\phi}) \hat{b}_{k\lambda\mu}].
\end{equation}
The form of this term stems from expanding the Hamiltonian in fluctuations around a homogeneous BEC of density $n$ and applying the Bogoliubov approximation and transformation (a constant mean-field shift is omitted). In such a case the Fourier-space interaction potentials, $U_\lambda(k)$, can be obtained in closed form~\cite{SchmidtLem15}:
\begin{equation}
\label{Ulambda}
U_\lambda(k)=\sqrt\frac{8nk^2\epsilon_k}{\omega_k(2\lambda+1)}\int dr 
r^2 V_\lambda(r)j_\lambda(kr),
\end{equation}
where $j_\lambda(kr)$ is the spherical Bessel function, and $V_\lambda(r)$ give 
the Legendre moments of the two-body interaction between the molecule and an 
atom from the BEC,  $V_\text{mol-at} (\mathbf{r}) = \sum_\lambda  V_\lambda (r) Y_{\lambda 0} (\theta_r, \phi_r)$ in the molecular frame. The spherical harmonic operators, $Y_{\lambda\mu}(\hat{\theta},\hat{\phi})$, in Eq.~\eqref{Himpbos} arise due to rotation of the molecule-atom interaction potential from the molecular to the laboratory frame. The explicit dependence on the molecule angle operators, 
$(\hat{\theta},\hat{\phi})$,  makes the angulon problem substantially different from other impurity problems such as the Bose 
polaron~\cite{Devreese15} and spin-boson~\cite{LeggettRMP87} models. It is important to note that while we consider the closed-form coupling of Eqs.~\eqref{Himpbos} and~\eqref{Ulambda} for simplicity, we expect the model to provide qualitative predictions beyond the range of applicability of the Bogoliubov approximation.  For other types of impurities, such as electrons or 
non-spherical atoms,  $U_\lambda(k)$ will assume a different form. Furthermore, 
the coupling constants $U_\lambda(k)$ are taken to be independent on the $\mu$ 
quantum number, which is the case, e.g.\ for linear molecular 
impurities~\cite{SchmidtLem15, LemSchmidtChapter}. Treating more complex, 
nonlinear molecules requires $\mu$-dependent potentials~\cite{Cherepanov}. 
However, the microscopic details of the impurity-bath interaction are not 
expected to alter the effects discussed in this paper qualitatively. Therefore, 
in what follows we use the coupling given by Eq.~\eqref{Ulambda}. 

\begin{figure}[h!]
\centering
\includegraphics[width=0.45\textwidth]{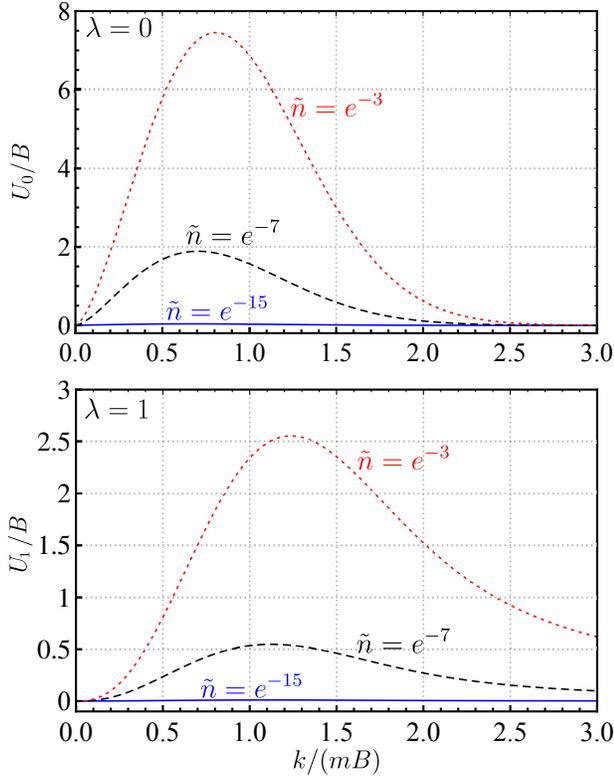}
\caption{
Dependence of the molecule-boson couplings $U_0$ and $U_1$ on $k$, for the parameters defined in Sec.~\ref{sec:zeeman}.  Each coefficient $U_\lambda(k)$ is shown at three 
different densities: $\tilde{n}=\exp(-15)$, $\tilde{n}=\exp(-7)$, $\tilde{n}=\exp(-3)$.  }
\label{fig:ulambda}
\end{figure}

Note that the long-wavelength behavior of Eq.~\eqref{Ulambda} is given by: 
\begin{equation}
 U_\lambda(k \to 0)\approx\zeta k^{\lambda+3/2}+\mathcal{O}(k^{\lambda+7/2}),
\end{equation}
where $\zeta$ is a constant independent of $\lambda$. The term 
$\zeta k^{\lambda+3/2}$ contributes to the rise of $U_\lambda(k)$ for small 
values of $k$, which is a consequence of the centrifugal barrier emerging for collisions with finite angular momentum. This behavior is illustrated  in Fig.~\ref{fig:ulambda}, where $U_\lambda(k)$ is plotted for $\lambda=0$ and $\lambda=1$ at several densities.

\section{The angulon self energy and spectral function}
\label{sec:variational}
In order to uncover the behavior of spinful angulons in a magnetic field, we 
make use of the equivalence between the variational and diagrammatic approaches 
to the angulon problem, see Refs.~\cite{SchmidtLem15, LemSchmidtChapter} for 
details. We start from the variational ansatz constructed of field-dependent 
molecular states and taking into account single-phonon excitations:
\begin{equation}
\label{ansatz}
\begin{split}
 \ket{\psi_{\tilde{J}LM_J}}&=
Z^{1/2}_{\tilde{J}LM_J}\ket{0}\ket{{\tilde{J},L,M_J}} \\ 
& +\sum\limits_{\substack{k\lambda \\ \tilde{j}lm_j\mu}}
\beta_{\lambda \tilde{j} l} (k)
C^{\tilde{J},M_J}_{\tilde{j},m_j;\lambda,\mu}\hat{b}^\dagger_{k\lambda\mu}
\ket{0}\ket{\tilde{j},l,m_j},
\end{split}
\end{equation}
where $\ket{0}$ is the vacuum of bosonic excitations,    and 
$Z^{1/2}_{\tilde{J}LM_J}$ and $\beta_{\lambda \tilde{j} l} (k)$ are the 
variational parameters obeying the following normalization condition:
\begin{equation}
 Z_{\tilde{J}LM_J}+\sum\limits_{k\lambda \tilde{j}l}|\beta_{\lambda 
\tilde{j}l}(k)|^2=1.
\end{equation}

Note that, despite the presence of the field, our variational coefficients are 
independent of $m_j$, which comes from the fact that the interaction potentials $U_\lambda (k)$ are 
independent of $\mu$.  Moreover,   in the presence of a field, $M_J$ is the only 
good quantum number of the system. In the variational ansatz of 
Eq.~\eqref{ansatz}, conservation of $M_J$ is accounted for by the Clebsch-Gordan 
coefficient 
$C^{\tilde{J},M_J}_{\tilde{j},m_j;\lambda,\mu}$~\cite{Varshalovich}. 
The number $\tilde{J}$, despite its presence in the Clebsch-Gordan coefficients 
is not a good quantum number since the ansatz is constructed on top of states 
which are not eigenstates of $\hat{\mathbf{J}}^2$. This means that in Eq.~\eqref{ansatz} we neglect 
the processes where $\tilde J$ changes due to the molecule-bath 
interactions, which is a good approximation away from crossings of levels with 
different $\tilde{J}$. This approximation becomes exact in the limit of $\eta\rightarrow0$, where $J$ is 
a good quantum number.

In Refs.~\cite{SchmidtLem15, LemSchmidtChapter} it has been shown, that by 
minimizing the functional
\begin{equation}
 F = \bra{\psi_{\tilde{J}LM_J}}\widehat{H}-E\ket{\psi_{\tilde{J}LM_J}}
 \end{equation}
over $Z_{\tilde{J}LM_J}$ and $\beta_{\lambda \tilde{j} l}(k)$, one can derive the following Dyson equation for the angulon: 
\begin{equation}
\label{Dyson}
G^{\textnormal{ang}}_{\tilde{J}LM_J}(E)^{-1}= G^0_{\tilde{J}LM_J}(E)^{-1}
-\Sigma_ {
\tilde{J}LM_J } (E),
\end{equation}
where $G^0_{\tilde{J}LM_J}(E)$ is the  free molecule Green's function
\begin{equation}
G^0_{\tilde{J}LM_J}(E)=\frac{1}{E^0_{\tilde{J}LM_J}-E},
\end{equation}
and 
\begin{equation}
\label{SelfE}
\begin{split}
 &\Sigma_{\tilde{J}LM_J}(E)=\\
 &=\sum\limits_{\substack{k \lambda \\  \tilde{j} l m_j m_j'}}
 \frac{
 U_\lambda(k)^2
 K^{\tilde{J}LM_J}_{\tilde{j}lm_j\lambda}(\eta) 
K^{\tilde{J}LM_J}_{\tilde{j}lm_j'\lambda}(\eta)}
 {\omega_k + 
\sum\limits_{m_j''}(C^{\tilde{J},M_J}_{\tilde{j},m_j'';\lambda,M_J-m_j''})^2
E^0_{\tilde{j}lm_j''}-E}
\end{split}
\end{equation}
is the angulon self energy. Here 
$K^{\tilde{J}LM_J}_{\tilde{j}lm_j\lambda}(\eta)$ is a coefficient resulting 
from the relevant angular momentum algebra, dependent upon Clebsch-Gordan 
coefficients, $a_{\tilde{J}LM_J}(\eta)$, and $b_{\tilde{J}LM_J}(\eta)$. The explicit 
form of 
$K^{\tilde{J}LM_J}_{\tilde{j}lm_j\lambda}(\eta)$ is given in 
Appendix~\ref{appendix:derivation}.  Furthermore, in the field-free limit, $\eta \to 0$, 
the expression for the self-energy~\eqref{SelfE} can be simplified, see  
Appendix~\ref{AppC} for a detailed derivation.

Thus, by casting the variational problem in terms of the Dyson equation~\eqref{Dyson}, we are able to access the energies of the excited states of the system by solving the following equation:
\begin{equation}
\label{dyson}
E=E^0_{\tilde{J}LM_J}-\Sigma_{\tilde{J}LM_J}(E),
\end{equation}
as well as to calculate the spectral function of the angulon, which is defined as:
\begin{equation}
\label{SpecFunc}
\mathcal{A}_{\tilde{J}LM_J}(E)= \textnormal{Im} [
G^ {
\textnormal { ang } } _ { \tilde{J}LM_J } (E+i0^+)].
\end{equation}

\begin{figure*}[t]
\centering
\includegraphics[width=0.85\textwidth]{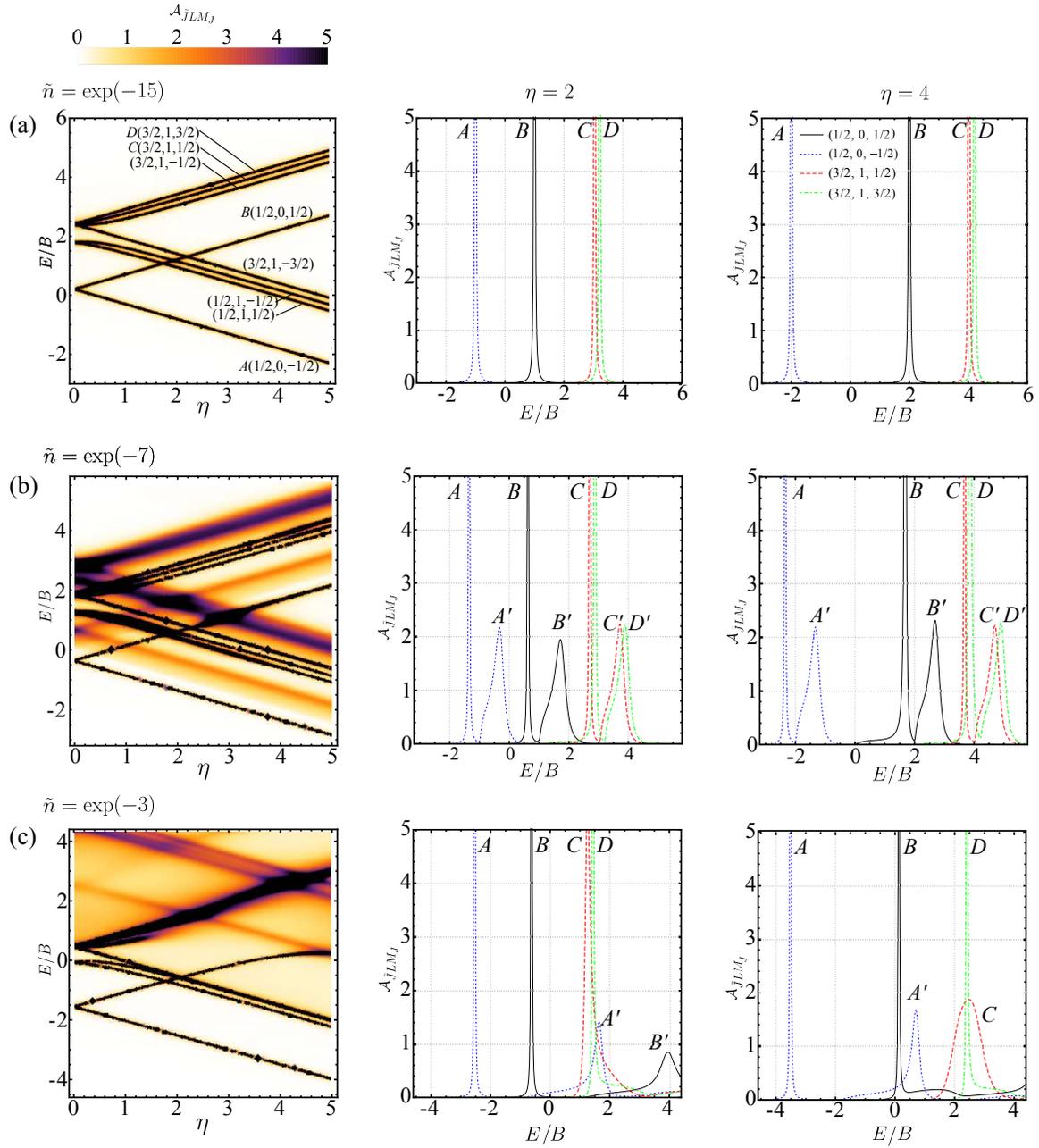}
\caption{
The angulon spectral function, 
$\mathcal{A}_{\tilde{J}LM_J}(E)$, for different dimensionless densities: 
(a)~$\tilde{n} = \exp(-15)$, which approximately corresponds to bare molecular 
states in a magnetic field as in Ref.~\cite{Friedrich2000}, (b) $\tilde{n} = 
\exp(-7)$, and (c) $\tilde{n} = \exp(-3)$. In the left column, the spectral 
functions are shown as a density plot, for all eight sublevels of $L=0$ and 
$L=1$, as a function of the magnetic field 
strength, $\eta$, and energy, $E/B$ (these states are shown one by one in Fig.~\ref{fig:spectral}). The middle and right columns show the energy dependence of the spectral function  for two selected values of $\eta=2$ and $\eta=4$ depicted for four states (1/2,0,-1/2), (1/2,0,1/2), (3/2,1,1/2), and (3/2,1,3/2) labeled as A, B, C, and D, respectively. The primed letters A', B', C', and D' mark additional fine structure emerging due to the interaction with the bath (see text).
}
\label{fig:spectraljoint}
\end{figure*}

\section{Zeeman effect for angulons}
\label{sec:zeeman}

In this section we study the angulon spectral function, Eq.~\eqref{SpecFunc}, in 
the presence of a magnetic field. In order to describe  the effects 
quantitatively, we use the Gaussian-shaped potentials, $V_\lambda(r)= u_\lambda 
(2\pi)^{-3/2}e^{-r^2/(2r_\lambda^2)}$ and the following parameter values in 
dimensionless units: $r_0=r_1=1.5 (mB)^{-1/2}$ and $a_{bb}=3.3(mB)^{-1/2}$, 
$u_0=1.75u_1=218B$, as previously used in Ref.~\cite{SchmidtLem15}.
Our choice of spin-rotation coupling is 
$\gamma=0.418B$, which is the value used in Ref.~\cite{Friedrich2000}.

In what follows, we focus on the substates belonging to $L=0$ and $L=1$ 
manifolds of rotational angular momentum.  Fig.~\ref{fig:spectraljoint} shows the 
dependence of the angulon spectral function on the field-strength parameter, 
$\eta$. Fig.~\ref{fig:spectraljoint}(a) corresponds to a vanishingly small 
density of the bath, and therefore 
reproduces the structure of bare molecular states in a magnetic 
field~\cite{Friedrich2000, LemFriPRA09}.

The $L=0$ and $L=1$ manifolds of rotational angular momentum contain eight bare 
molecular Zeeman levels 
$(\tilde{J} , L, M_J)$ with their parity given by $P=(-1)^L$. As the 
magnetic field couples only levels with the same $L$, the parity remains 
unchanged in the presence of the field. For extreme 
values of the projection $M_J$, i.e. $M_J=\pm\tilde{J}$ with $\tilde{J}=L+1/2$, 
the field-dependence 
of the levels, as given by Eqs.~\eqref{eigenenergy}--\eqref{eigenenergy2},
reduces to the linear one. For other states, linearity of the energy dependence on 
the field occurs in the high-field regime due to Paschen-Back uncoupling of 
spin from rotation of the molecule.

Figures~\ref{fig:spectraljoint}(b) and (c) reveal that for a finite bath density,  the molecular levels are  shifted towards lower energies.  This effect, 
known as polaron shift, is a result of isotropic interactions between the 
impurity and the bath and has been widely studied for structureless 
impurities~\cite{Devreese15}. Apart from the polaron shift, 
Figures~\ref{fig:spectraljoint}(b) and (c) reveal a complex spectral structure 
emerging from the molecule-bath interaction. 
Namely, a lot of metastable states (shades of yellow) appear in between the stable angulon states (dark lines).

In order to understand this fine structure in detail, in the middle and right columns of Fig.~\ref{fig:spectraljoint} we present the spectral functions for the four selected states, $A (1/2, 0, -1/2)$, $B (1/2, 0, 1/2)$, $C (3/2, 1, 1/2)$, and $D (3/2, 1, 3/2)$, at two selected values of the magnetic field strength, $\eta=2$ (middle column) and $\eta=4$ (right column).

\begin{figure}[!ht]
\centering
\includegraphics[width=0.46\textwidth]{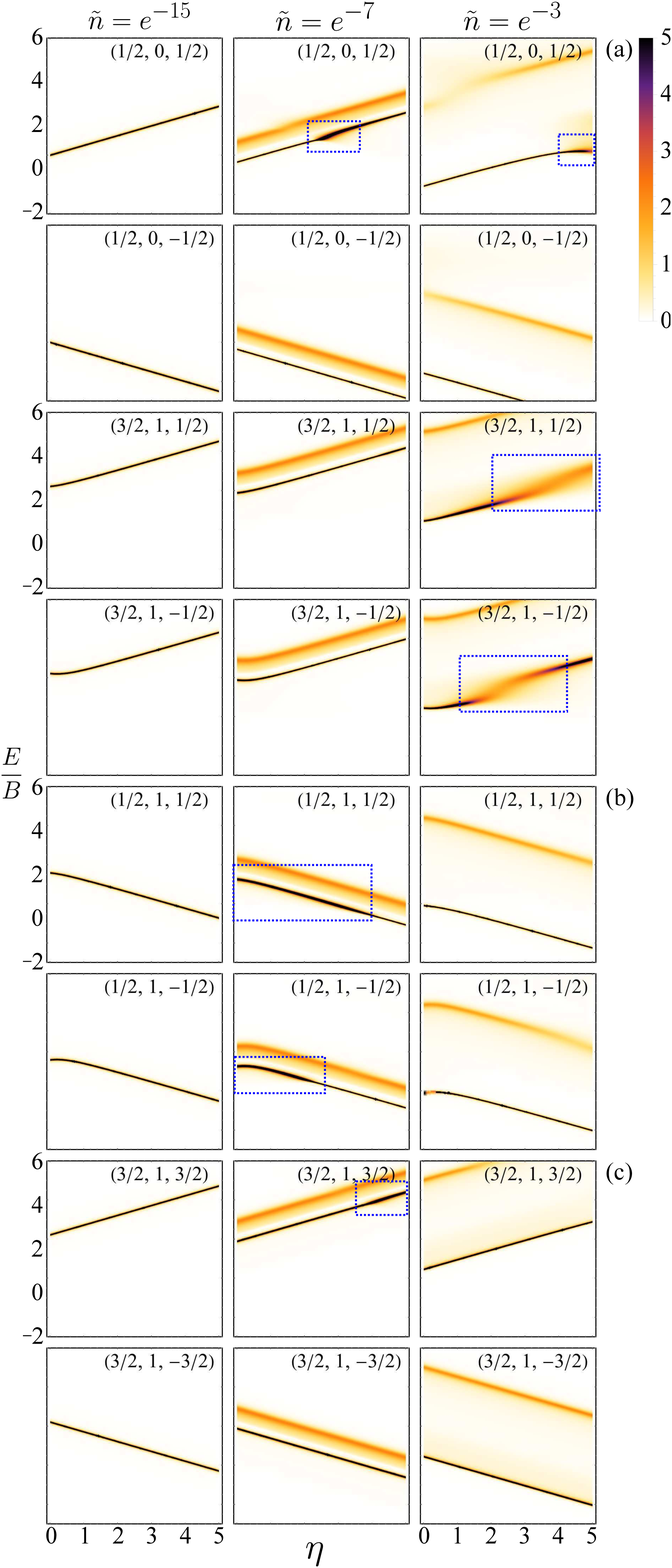}
\caption{ 
Spectral functions of all eight $L=0$ and $L=1$ states vs. magnetic field 
strength and energy for densities $\tilde{n} = \exp(-15)$, $\tilde{n} = \exp(-7)$, $\tilde{n} = \exp(-3)$. The 
first column on the left (density of $\tilde{n} = \exp(-15)$) corresponds to 
bare molecular states in magnetic field as 
in Ref.~\cite{Friedrich2000} The labels (a), (b), and (c) on the right correspond to panels in Fig.~\ref{fig:beta}. The blue dotted frames mark clearly visible angulon instabilities.
}
\label{fig:spectral}
\end{figure}

The observed features are qualitatively similar to that predicted in 
Ref.~\cite{SchmidtLem15} for the spinless angulon in the absence of a magnetic 
field. For the vanishingly low density of the bath, $\tilde{n}=\exp(-15)$, 
Fig.~\ref{fig:spectraljoint}(a), each of the states is given by a sharp peak, 
which approximately coincides with the molecule spectrum in the absence of a 
bath. For a finite density, $\tilde{n}=\exp(-7)$, an additional fine structure 
emerges in the spectrum, as labeled by the primed letters $A'$, $B'$, $C'$, and 
$D'$. This is the so-called Many-Body-Induced Fine Structure (MBIFS) of the 
first kind~\cite{SchmidtLem15}, which emerges due to dressing of the stable 
angulon state with a phonon excitation carrying zero angular momentum, 
$\lambda=0$ -- this effect is described in more detail below. In a magnetic 
field, the position of this phonon wing changes in the same way as that of the 
stable angulon state: the states $B$, $C$, and $D$ are shifted towards higher 
energies for larger $\eta$, while the energy of the $A$ state decreases. For an 
even larger density, $\tilde{n}=\exp(-3)$, Fig.~\ref{fig:spectraljoint}(c), the 
splitting between the stable angulon peak and the attached phonon continuum 
increases further. Note that the features $C'$ and $D'$ for $\eta=2$ and $B'$, 
$C'$, and $D'$ for $\eta=4$ move outside the range  of the corresponding 
plots. Moreover, the further the phonon branch is from the main angulon line, 
the broader is the spectral feature associated with it.

\begin{figure}[!ht]
\centering
\includegraphics[width=0.49\textwidth]{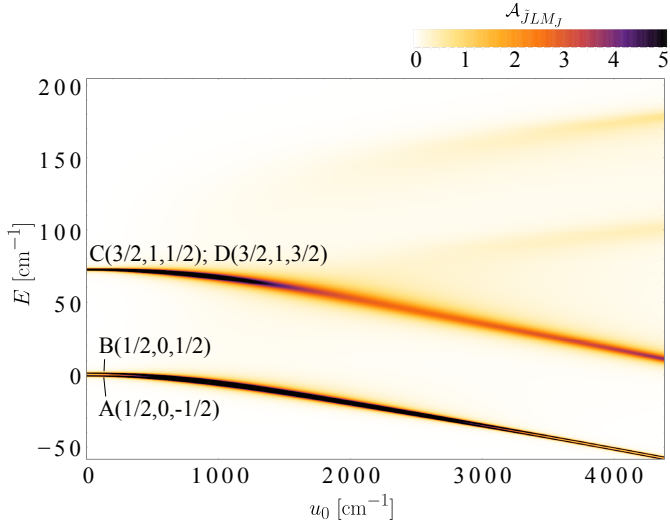}
\caption{ 
{ An example of a spectral function in absolute units, plotted  for a molecule with $B = 14.603$ cm$^{-1}$ (that of CH), spin-rotation coupling $\gamma =0.4B$, in a field of $\mathscr{H} = 2$~T, as a function of the effective molecule-helium interaction $u_0$ and the energy $E$. The molecule is immersed in superfluid $^4$He (superfluid density $n = 10^{22}$~cm$^{-3}$).  The states $C$ and $D$ are labeled jointly as they lie too close to be resolved. }}
%%\btext{[FILL IN DETAILS]}}
\label{fig:absunitsplot}
\end{figure}

To provide a detailed analysis of the physical 
phenomena happening in the presence of the bath and a magnetic field, in 
Fig.~\ref{fig:spectral} we 
study the states one by one. The left column presents bath-free bare molecular
states. While moving to the middle and right columns, first we notice the 
splitting of the levels due to the 
MBIFS which leads to splitting of lines in each plot into a doublet. This splitting results from the isotropic term $U_0(k)$ 
of the molecule-bath interaction in Eq.~\eqref{Himpbos} and can be 
understood (approximately) as a splitting between the states 
$\ket{\tilde{J},L,M_J}\otimes\ket{\textnormal{no phonons}}$ and 
$\ket{\tilde{J},L,M_J}\otimes\ket{\textnormal{one
phonon with }\lambda=0}$. This effect is reminiscent of the phonon wings predicted for acetylene molecules in He nanodroplets in Ref.~\cite{ZillichHeC2H2}, and will not be the main focus of our studies.

Secondly, we observe the
emergence of instabilities which results from 
anisotropic interactions, $U_1(k)$ of Eq.~\eqref{Himpbos}. The instabilities are 
highlighted in Fig.~\ref{fig:spectral} by the blue dotted frames. It has been 
previously shown that such instabilities are accompanied by the transfer of  
angular momentum  from the impurity to the bath, i.e.\ the resonant emission of 
phonons with nonzero angular momentum. Since in our model we include only the 
leading anisotropic interaction term, $U_1(k)$, the emitted phonon carries 
angular momentum of $\lambda=1$, however excitations with higher angular 
momentum are also possible~\cite{Redchenko16}. As an example, the state $(1/2, 
0, 1/2)$ features the angulon  instabilities around $\eta=2.8$ at the density of 
$\tilde{n}=\exp(-7)$ and around $\eta=4.8$ at the density $\tilde{n}=\exp(-3)$, 
as can be seen from the first row of Fig.~\ref{fig:spectral}. These 
instabilities occur due to the interaction with the phonon continua of the states (1/2,1,1/2) and (1/2,0,-1/2), respectively.
Another example is an instability of state $(3/2, 
1, -1/2)$, taking place around $\eta=2.5$ at the density of $\tilde{n}=\exp(-3)$, which is due to interaction with the phonon continuum of the state $(1/2, 0, -1/2)$.  
It was previously shown that the instabilities discussed above  lead to 
anomalous screening of the impurities' dipole moments and 
polarizabilities~\cite{Yakaboylu16}, and can be manipulated using an  external 
electrostatic field~\cite{Redchenko16}.  
Crucially, from Fig.~\ref{fig:spectral} one can see that the position of the instabilities depends on the magnitude of the magnetic field $\eta$ as well. This paves the way to control the resonant emission of phonons with a given angular momentum, as discussed in the following section. {As an example, in Fig.~\ref{fig:absunitsplot}, plotted in absolute units, we illustrate the emergence of angulon instabilities for  a molecule with the rotational constant  $B = 14.603$ cm$^{-1}$ (equal to that of CH), and the spin-rotation coupling  $\gamma=0.4B$, in a field of $\mathscr{H} = 2$~T. Although Fig.~\ref{fig:absunitsplot} is plotted for a high solvent density of $n = 10^{22}$ cm$^{-3}$, we also quantitatively account for lower interactions (such as those occurring at lower densities of weakly-interacting  BEC's) by including the regime of small interaction parameter $u_0$ in the plot (the ratio of $u_0/u_1 = 1.75$ is kept constant).}

\section{Angular momentum of phonons in the solvent}
\label{sec:manipulating}
In this section we study the dependence of the phonon populations in 
individual $\lambda$ channels, 
\begin{equation}
\label{totaldensities}
\tilde \beta_\lambda(k)=\sum\limits_{\tilde{j}l}|\tilde{\beta}_{\lambda\tilde{j}l}(k)|^2,
\end{equation}
on the magnitude of the magnetic field. Here, the 
phonon populations $\tilde{\beta}_{\lambda\tilde{j}l}(k)$ can be obtained from the variational calculations as follows:
\begin{equation}
\label{eq:beta}
\tilde{\beta}_{\lambda 
\tilde{j}l}(k)=-\frac{
U_\lambda(k)
 K^{\tilde{J}LM_J}_{\tilde{j}lm_j\lambda} 
}{\sum\limits_{m_j''}(C^{J,M_J}_{j,m_j'';\lambda,M_J-m_j''}
)^2E^0_{jlm_j''}-E+\omega_k},
\end{equation}
and the energy $E$ found as a solution of Eq.~\eqref{dyson}. The phonon populations $\tilde{\beta}_{\lambda\tilde{j}l}(k)$ are related to the variational coefficients of Eq.~(\ref{ansatz}) by the following normalization relation:
\begin{equation}
\label{eq:normalization}
\beta_{\lambda \tilde{j}l}(k)=\frac{\tilde{\beta}_{\lambda 
\tilde{j}l}(k)}{\left(1+\sum\limits_{k\lambda \tilde{j}l}|\tilde{\beta}_{\lambda 
\tilde{j}l}(k)|^2\right)^{1/2}}.
\end{equation}
We elaborate on the mathematical difficulties of normalization in Appendix~\ref{Appbeta} and discuss the unnormalized quantities as they fully show the proportion between phonon populations for different $\lambda$ channels. 

\begin{figure}[h!]
\centering
\includegraphics[width=0.48\textwidth]{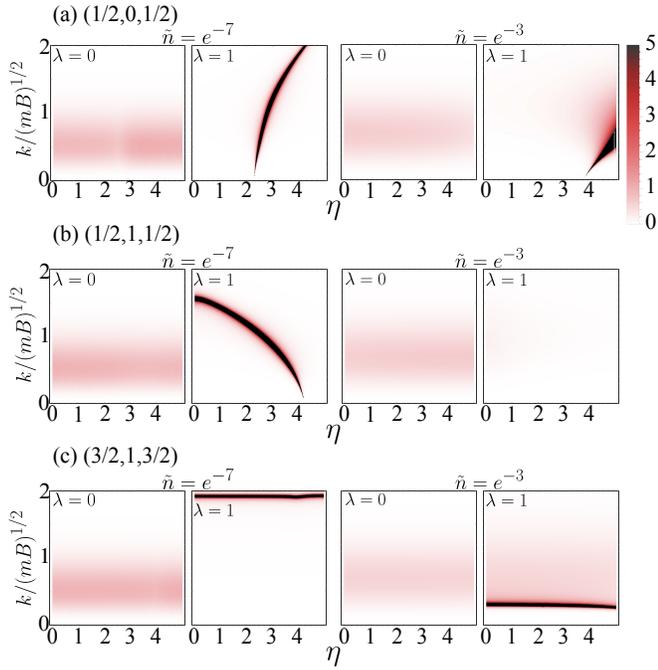}
\caption{
Dependence of the phonon populations $\tilde \beta_\lambda(k)$ on the magnitude of the magnetic field $\eta$ and wavevector $k$ for two first channels $\lambda=0,1$ in two different densities $\tilde n=e^{-7}$ and $\tilde n =e^{-3}$. The three panels feature different states: (a) (1/2,0,1/2), (b) (1/2,1,1/2), (c) (3/2,1,3/2).}
\label{fig:beta}
\end{figure}

Fig.~\ref{fig:beta} shows the phonon populations $\tilde \beta_\lambda(k)$ for two lowest channels, $\lambda=0,1$, depending on the magnetic field strength $\eta$ and the wavevector $k$. As discussed above, the phonons with $\lambda=0$ emerge as a phonon wing surrounding the main angulon state, while the $\lambda=1$ phonons are resonantly emitted at the angulon instability. This reflects itself in moderate emission of phonons into the $\lambda=0$ channel across all studied densities and magnetic field and infinitely growing $\tilde \beta_1(k)$ at the instabilities. By comparing Figs.~\ref{fig:spectral} and~\ref{fig:beta} we can see that the strong emissions in $\lambda=1$ channel occur for the same range of the field as the instabilities in the spectral function. Let us note that the instabilities can look differently on the spectral function plots -- they can be manifested as line discontinuities or line broadening. Moreover, stable shape of the spectral function of (1/2,1,1/2) state for $\tilde n=e^{-3}$ corresponds to almost no emission to $\lambda=1$ for this state and density.

Thus, by manipulating the angulon energies with a magnetic field, one can control the emission of phonons with a given angular momentum $\lambda$, and fine-tune the  phonon populations in different $\lambda$-channels.

\section{Conclusions}
\label{sec:discussion}

In summary, we used the angulon theory to study  a quantum impurity with 
spin$-1/2$ and rotational  angular momentum, immersed in a many-particle bath of 
bosons, in the presence of an external magnetic field. We have shown that the 
field can be used to manipulate the exchange of  angular momentum between 
the impurity and the bath, as mediated by spin-rotation coupling. In turn, this 
paves the way to manipulate the positions of the angulon 
instabilities~\cite{SchmidtLem15, SchmidtLem16, LemSchmidtChapter, Yakaboylu16, 
Redchenko16}, and thereby control the angular momentum of phonons in the bath. Recently, 
the signatures of the angulon instabilities were found in spectra of molecules 
in helium nanodroplets~\cite{Cherepanov}, which opens up a possibility to test 
the presented predictions in experiment. In particular, we expect that the effects predicted in this paper can be detected through electron spin resonance measurements on molecules in superfluid helium nanodroplets~\cite{KochPRL09}. {For  typical  $^2\Sigma$ molecules, 
such as CaF, SrF, and CH, the respective rotational constants are~0.338~
cm$^{-1}$~\cite{Field1975107}, 0.250~cm$^{-1}$~\cite{Domaille1977}, and 14.603~cm$^{-1}$~\cite{Constants}  which 
makes $\eta=1$ correspond to a field of $0.36$~T, $0.27$~T,  and $15.62$~T, respectively.} 
{Table~\ref{quantititestable} lists the quantities used throughout the paper in absolute units for   $^2\Sigma$ electronic ground  states of CaF, SrF, and CH molecules. 
Substantially smaller field magnitudes are expected to be required for molecules containing
highly magnetic elements, such as erbium~\cite{AikawaPRL12} and 
dysprosium~\cite{LuPRL11}. We note that in the present work translational motion of molecules in a superfluid has been neglected. The rotation-translation coupling can lead to additional inhomogeneous broadening of the spectroscopic lines \cite{LehmannMolPhys99} and we are currently extending the angulon model to account for it. }
 
Finally, we would like to note that the formalism presented here is quite general, and 
can be applied to any spinful impurity possessing rotational or orbital angular momentum, 
immersed into, in principle, any kind of a bosonic bath. Therefore, the 
predicted effects  can find  potential applications not only for 
molecules trapped in superfluid helium nanodroplets~\cite{ToenniesAngChem04} 
and ultracold gases~\cite{Midya2016}, but also to Rydberg excitations in 
Bose-Einstein condensates~\cite{Balewski:2013hv} and non-equilibrium magnetism 
in solids~\cite{KoopmansPRL05, ChudnovskyPRB05, Zhang2014, Tows2015, 
Garanin2015, Tsatsoulis2016}.

\begin{table}[H]
\centering
\caption{{ Values of energy, bath density, and magnetic field in absolute units for CaF, SrF, and CH molecules immersed in superfluid~$^4$He.}}
\label{quantititestable}
\vspace{0.1cm}
\begin{tabular}{c|cc|cc|cc}
Molecule & \multicolumn{2}{c|}{CaF} & \multicolumn{2}{c|}{SrF} & \multicolumn{2}{c}{CH}  \\ \hline
E=1 & \multicolumn{2}{c|}{$0.338$ cm$^{-1}$} & \multicolumn{2}{c|}{$0.250$ cm$^{-1}$} & \multicolumn{2}{c}{$14.603$ cm$^{-1}$} \ \\[3pt]
$\tilde n=\exp(-15)$ & \multicolumn{2}{c|}{$2.46 \cdot 10^{13}$ cm$^{-3}$} & \multicolumn{2}{c|}{$1.56 \cdot 10^{13}$ cm$^{-3}$} & \multicolumn{2}{c}{$6.98 \cdot 10^{15}$ cm$^{-3}$} \\
$\tilde n=\exp(-7)$ &\multicolumn{2}{c|}{$7.32 \cdot  10^{16}$ cm$^{-3}$} &\multicolumn{2}{c|}{$4.66 \cdot 10^{16}$ cm$^{-3}$} &\multicolumn{2}{c}{$2.08 \cdot 10^{19}$ cm$^{-3}$}\\
$\tilde n=\exp(-3)$ & \multicolumn{2}{c|}{$4.00 \cdot  10^{18}$ cm$^{-3}$} & \multicolumn{2}{c|}{$2.54 \cdot 10^{18}$ cm$^{-3}$} & \multicolumn{2}{c}{$1.14 \cdot 10^{21}$ cm$^{-3}$}\\[3pt]
$\eta=1$ & \multicolumn{2}{c|}{0.36 T} & \multicolumn{2}{c|}{0.27 T} & \multicolumn{2}{c}{15.62 T} \\
\end{tabular}
\vspace{0.1cm}
\end{table}

\acknowledgments{We acknowledge insightful discussions with Giacomo Bighin, Igor Cherepanov, Johan Mentink, and Enderalp Yakaboylu. This work was supported by the Austrian Science Fund (FWF), project Nr. P29902-N27. W.R. was supported by the Polish Ministry of Science and Higher Education grant 
number MNISW/2016/DIR/285/NN and by the European Union's Horizon 2020 research and innovation programme under the Marie Sk{\l}odowska-Curie Grant Agreement No. 665385.

\onecolumngrid

\newpage
\appendix
\section{Details of the impurity-field interaction}
\label{appendix_details}
In this Appendix, we calculate the coefficients $a_{\tilde{J}LM_J}(\eta)$ and 
$b_{\tilde{J}LM_J}(\eta)$ resulting from diagonalization of the 
molecular and magnetic 
part 
of the 
Hamiltonian written in the field-independent basis: $\ket{J=L+1/2,L,M_J}$, 
$\ket{J=L-1/2,L,M_J}$. 
For given $L,M_J$, the molecular and magnetic part of the Hamiltonian in the matrix 
form 
reads:

\begin{equation}
\widehat{H}_{\textnormal{mol}}+
\widehat{H}_{\textnormal{mol-f}}=\left[
\begin{array}{cc}
 BL (L+1)+\frac{L \gamma }{2}+\frac{B\eta M_J }{ 2\left(L+\frac{1}{2}\right)} & 
-\frac{B\eta}{2} \sqrt{1-\frac{M_J^2}{\left(L+\frac{1}{2}\right)^2}}   \\
 -\frac{B\eta}{2} \sqrt{1-\frac{M_J^2}{\left(L+\frac{1}{2}\right)^2}}   & BL 
(L+1)-\frac{\gamma}{2}  (L+1)-\frac{B\eta M_J }{ 2\left(L+\frac{1}{2}\right)}
\\
\end{array}
\right].
\end{equation}

Upon diagonalization we find eigenenergies as given by Eq.~\eqref{eigenenergy} and corresponding eigenstate coefficients given by:
\begin{equation}
a_{\tilde JLM_J}(\eta)=
\frac{-\phi_1+2(\tilde J-L)\sqrt{\phi_1^2+\phi_2^2}}{\sqrt{\phi_2^2+(-\phi_1+2(\tilde J-L)\sqrt{\phi_1^2+\phi_2^2})^2}}
,
\end{equation}
and
\begin{equation}
b_{\tilde JLM_J}(\eta)=
\frac{\phi_2}{\sqrt{\phi_2^2+(-\phi_1+2(\tilde J-L)\sqrt{\phi_1^2+\phi_2^2})^2}}
,
\end{equation}
where 
\begin{equation}
\phi_1=\frac{B\eta M_J}{L+\frac{1}{2}}+\gamma\left(L+\frac{1}{2}\right),\quad\phi_2=B\eta\sqrt{1-\frac{M_J^2}{\left(L+\frac{1}{2}\right)^2}}. 
\end{equation}

\section{Detailed derivation of the Dyson equation}
\label{appendix:derivation}
We denote the terms of the ansatz from Eq.~\eqref{ansatz}  as 
$\ket{\psi_{\tilde{J}LM_J}}=\ket{\psi_{\tilde{J}LM_J}^1}+\ket{\psi_{\tilde{J}
LM_J}^2}$, where:
\begin{equation}
\ket{\psi_{\tilde{J}LM_J}^1}=Z^{1/2}_{\tilde{J}LM_J}\ket{0}\ket{{\tilde{J},L,
M_J}} ,
\end{equation}
and
\begin{equation}
\ket{\psi_{\tilde{J}LM_J}^2}=\sum\limits_{\substack{k\lambda \\\tilde{j} 
lm_j\mu}}
\beta_{\lambda\tilde{j}l}(k)
C^{\tilde{J},M_J}_{\tilde{j},m_j;\lambda,\mu}
\hat{b}^\dagger_{k\lambda\mu}\ket{0}\ket{\tilde{j},l,m_j}.
\end{equation}
We minimize the functional 

\begin{equation}
\begin{split}
F&=\bra{\psi_{\tilde{J}LM_J}}\widehat{H}\ket{\psi_{\tilde{J}LM_J}}-E\braket{
\psi_{\tilde{J}LM_J}}{ \psi_{\tilde{J}LM_J}}\\
&=\bra{\psi_{\tilde{J}LM_J}^1}\widehat{H}\ket{\psi_{\tilde{J}LM_J}^1}+\bra{
\psi_{\tilde{J}LM_J}^2}\widehat{H} \ket{\psi_{\tilde{J}LM_J}^2}
+\left[
\bra { \psi_{\tilde{J}LM_J}^1 } \widehat{H}\ket { \psi_{\tilde{J}LM_J}^2 }
+c.c.\right]-E\braket{\psi_{\tilde{J}LM_J}}{\psi_{\tilde{J}LM_J}}.
\end{split}
\end{equation}
To calculate $F$, we sequentially evaluate its 
terms. First,
\begin{equation}
\label{psipsi}
\braket{\psi_{\tilde{J}LM_J}}{\psi_{\tilde{J}LM_J}}=|Z_{\tilde{J}LM_J}|+
\sum\limits_{\substack{k\lambda \\\tilde{j} l}}
|\beta_{\lambda \tilde{j}l}(k)|^2,
\end{equation}
where we summed over $m_j$ and $\mu$ using the orthogonality relations for Clebsch-Gordan 
coefficients~\cite{Varshalovich}. Then 
\begin{equation}
\label{psi1Hpsi1}
\bra{\psi_{\tilde{J}LM_J}^1}\widehat{H}\ket{\psi_{\tilde{J}LM_J}^1}=
|Z_{\tilde{J}LM_J}|E^0_{\tilde{J}LM_J},
\end{equation}
and
\begin{equation}
\label{psi2Hpsi2}
\bra{\psi_{\tilde{J}LM_J}^2}\widehat{H}\ket{\psi_{\tilde{J}LM_J}^2}=
\sum\limits_{k \lambda \tilde{j} l }
|\beta_{ \lambda  \tilde{j} l }(k)|^2
\left[
\sum\limits_{m_j}(C^{\tilde{J},M_J}_{\tilde{j},m_j;\lambda,M_J-m_j})^2E^0_{\tilde{j}lm_j}+\omega_k
\right].
\end{equation}
The contribution of the interaction (fourth term of Eq.~\eqref{hamiltonian}) vanishes in Eqs.~\eqref{psipsi}, \eqref{psi1Hpsi1}, \eqref{psi2Hpsi2}, unlike the 
$\bra{\psi_{\tilde{J}LM_J}^1}\widehat{H}\ket{\psi_{\tilde{J}LM_J}^2}$ term, 
which is nonzero solely due to the 
interaction. We express the field-dependent states in the basis of Eq.~\eqref{superposition}, and decompose these states further as in Eq.~(\ref{barestates}). Now we can act with the spherical harmonics operators on the 
rotational kets, with the matrix elements given by~\cite{Varshalovich, LemSchmidtChapter}:
\begin{equation} 
 \bra{L',M'}Y_{\lambda\mu}(\hat{\theta},\hat{\phi})\ket{L,M}
=\sqrt{\frac{
(2L+1)(2\lambda+1) } { 4\pi(2L'+1) } 
}C^{L'M'}_{LM;\lambda\mu}C^{L'0}_{L0;\lambda0}.
\end{equation}
Then, making use of the spin kets orthogonality we arrive at the final form of 
$\bra{\psi_{\tilde{J}LM_J}^1}\widehat{H}\ket{\psi_{\tilde{J}LM_J}^2}$:

\begin{equation}
\bra{\psi_{\tilde{J}LM_J}^1}\widehat{H}\ket{\psi_{\tilde{J}LM_J}^2}=
(Z_{ \tilde{J} L M_J}^*)^{1/2}
\sum\limits_{k \lambda   \tilde{j} l m_j}
\beta_{ \lambda \tilde{j} l  }(k)
U_\lambda(k)
K^{\tilde{J}LM_J}_{\tilde{j}lm_j\lambda}(\eta),
\end{equation}
where we denoted

\begin{equation}
\begin{split}
&K^{\tilde{J}LM_J}_{\tilde{j}lm_j\lambda}(\eta)=
C^{\tilde J,M_J}_{\tilde j,m_j;\lambda,M_J-m_j}
\sqrt{\frac{(2\lambda+1)}{4\pi}}(-1)^\lambda C^{l0}_{L0\lambda0}
\\
\times&\left\{
\left[C^{L+\frac{1}{2},M_J}_{L,M_J-\frac{1}{2};\frac{1}{2},\frac{1}{2}}a_{\tilde
{ J } LM_J
}(\eta)+C^{L-\frac{1}{2},M_J}_{L,M_J-\frac{1}{2};\frac{1}{2}
, \frac{1}{2} } b_ {
\tilde{J}LM_J}(\eta)\right]\left[C^{l+\frac{1}{2},m_j}_{l,m_j-\frac{1}{2};\frac{
1 } { 2
} ,
\frac{1}{2}}a_{\tilde{j}lm_j}
(\eta)+C^ { l-\frac{1}{2} , m_j } _ { l ,
m_j-\frac{1}{2};\frac{1}{2},\frac{1}{2}}b_{\tilde{j}lm_j}(\eta)\right]C^{L,
M_J-\frac { 1
}{2}}_{l, m_j-\frac{1}{2};\lambda , M_J-m_j 
}+\right.\\
&\left.\left[C^{L+\frac{1}{2},M_J}_{L,M_J+\frac{1}{2};\frac{1}{2},-\frac{1}{2}}
a_ {
\tilde{J}LM_J}(\eta)+C^{L-\frac{1}{2},M_J}_{L,
M_J+\frac{1}{2};\frac{1}{2} , -\frac{1}{2} } b_ {
\tilde{J}LM_J}(\eta)\right]\left[C^{l+\frac{1}{2},m_j}_{l,m_j+\frac{1}{2};\frac{1}{2
} ,
-\frac{1}{2}}a_{\tilde{j}lm_j}
(\eta)+C^ { l-\frac{1}{2} , m_j } _ { l ,
m_j+\frac{1}{2};\frac{1}{2},-\frac{1}{2}}b_{\tilde{j}lm_j}(\eta)\right]C^{L,
M_j+\frac {
1}{2}}_{l, m_j+\frac{1}{2};\lambda , M_J-m_j 
} 
 \right\}.
 \end{split}
\end{equation}
with $a_{\tilde{J}LM_J}(\eta)$ and $b_{\tilde{J}LM_J}(\eta)$ calculated in 
Appendix~\ref{appendix_details}.

Having calculated the functional $F$, we can now compute its derivatives with 
respect to the variational parameters:
\begin{equation}
\label{der1}
 \frac{\partial F}{\partial (Z_{\tilde{J}LM_J}^*)^{1/2}}=
 (Z_{\tilde{J}LM_J})^{1/2}(E^0_{\tilde{J}LM_J}-E)+\sum\limits_{k\lambda 
\tilde{j}lm_j}\beta_{\lambda 
\tilde{j}l}(k) U_\lambda(k)K^{\tilde{J}LM_J}_{\tilde{j}lm_j\lambda} \equiv 0,
\end{equation}

\begin{equation}
\label{der2}
 \frac{\partial F}{\partial \beta_{\lambda \tilde{j}l}^*(k)}=
\beta_{ \lambda \tilde{j} l  }(k)
\left[ 
 \sum\limits_{ m_j}
(C^{\tilde{J},M_J}_{\tilde{j},m_j;\lambda,M_J-m_j})^2E^0_{\tilde{j}lm_j}-E+\omega_k
\right]
+
(Z_{\tilde{J}LM_J})^{1/2}U_\lambda(k)\sum\limits_{m_j}K^{\tilde{J}LM_J}_{\tilde{j}lm_j\lambda} \equiv 0.
\end{equation}

Then, upon substitution of $\beta_{\lambda \tilde{j}l }(k)$ from Eq.~\eqref{der2} 
into Eq.~\eqref{der1}, $Z_{\tilde{J}LM_J}^{1/2}$ cancels out and we arrive at 
the 
Dyson equation as given by Eq.~\eqref{dyson}.

\section{The field-free case}
\label{AppC}

In this Appendix, we describe an intermediate case between that discussed in 
the main text and the original angulon theory introduced in 
Ref.~\cite{SchmidtLem15}. Namely, we consider a  zero-field limit, $\eta=0$, in the presence of the 
spin-rotation coupling, $\gamma\neq0$.

The new variational ansatz reads:
\begin{equation}
\label{nonmagneticansatz}
\ket{\psi_{JLM_J}}=
\underbrace{Z^{1/2}_{JLM_J}\ket{0}\ket{J,L,M_J}}_{:=\ket{
\psi_{JLM_J}^1 }} + 
 \underbrace{\sum\limits_{\substack{k\lambda\mu \\ jl m_j}}
\beta_{\lambda jl}(k)
C^{J,M_J}_{j,m_j;\lambda,\mu}
\hat{b}^\dagger_{k\lambda\mu}\ket{0}\ket{j,l,m_j}}_{:=\ket{
\psi_{JLM_J}^2 }}.
\end{equation}
As in the case when the field is present, we minimize the functional 

\begin{equation}
\begin{split}
F&=\bra{\psi_{JLM_J}}\widehat{H}\ket{\psi_{JLM_J}}-E\braket{\psi_{JLM_J}}{\psi_{
JLM_J}} \\
&=\bra{\psi_{JLM_J}^1}\widehat{H}\ket{\psi_{JLM_J}^1}
+\bra{\psi_{JLM_J}^2
}\widehat{H}\ket {\psi_{JLM_J}^2}
 +(\bra{
\psi_{JLM_J}^1} \widehat{H}\ket{\psi_{JLM_J}^2}
+c.c)-E\braket{\psi_{JLM_J}}{\psi_{JLM_J}}.
\end{split}
\end{equation}
We calculate $F$ term-by-term:
\begin{equation}
\braket{\psi_{JLM_J}}{\psi_{JLM_J}}=
|Z_{JLM_J}|
+\sum\limits_{k \lambda j l}
|\beta_{ \lambda j l}(k)|^2,
\end{equation}
and
\begin{equation}
\bra{\psi_{JLM_J}^1}\widehat{H}\ket{\psi_{JLM_J}^1}=
|Z_{JLM_J}|
E^0_{JL}.
\end{equation}
where $E^0_{JL}$
is the energy of the $\ket{J,L,M_J}$ state as defined by 
Eq.~(\ref{bareenergies}).

The next term reads
\begin{equation}
\bra{\psi_{JLM_J}^2}\widehat{H}\ket{\psi_{JLM_J}^2}=
\sum\limits_{k \lambda   j l }
|\beta_{ \lambda j l}(k)|^2
\left(\omega_k + E^0_{jl}\right).
\end{equation}

Furthermore, we have $\bra{\psi_{JLM_J}^1}\widehat{H}\ket{\psi_{JLM_J}^2}=\bra{\psi_{JLM_J}^1}
\widehat {H}_{ \textnormal{mol-bos}} \ket{
\psi_{JLM_J}^2 } $, from which we obtain:
\begin{equation}
\bra{\psi_{JLM_J}^1}\widehat{H}\ket{\psi_{JLM_J}^2}=
(Z^*_{JLM_J})^{1/2}\sum\limits_{k \lambda j l }U_\lambda(k)
\beta_{\lambda j l }(k)K^{JL}_{jl\lambda},
\end{equation}
where

\begin{equation}
K^{JL}_{jl\lambda}=C^{L,0}_{l,0;\lambda,0}
\sqrt{\frac{(2\lambda+1)(2l+1)}{4\pi}}
\times 
4(J-L)(j-l)(-1)^{1/2+l+\lambda+J}
\sqrt{(2(l+J-L)+1) } 
\left\{
\begin{array}{ccc}
 1/2 & l & j \\
 \lambda & J & L
\end{array}
\right\}.
\end{equation}

Now we vary $F$ with respect to the parameters, obtaining:
\begin{equation}
\label{first::fieldless}
\frac{\partial F}{\partial (Z_{JLM_J}^{1/2})^*}=
Z_{JLM_J}^{1/2}
(E^0_{JL}-E)
+\sum\limits_{k \lambda  jl }U_\lambda(k)K^{JL}_{jl\lambda}\equiv 0,
\end{equation}
\begin{equation}
\label{second::fieldless}
\frac{\partial F}{\partial \beta_{ \lambda jl}(k)^*}=
\beta_{ \lambda jl }(k)
(\omega_k + E^0_{jl}-E)+
Z_{JLM_J}^{1/2}U_\lambda(k)
K^{JL}_{jl\lambda}\equiv 0.
\end{equation}
Then, substituting $\beta$ from Eq.~\eqref{second::fieldless} into 
Eq.~\eqref{first::fieldless} we arrive at the following Dyson equation:
\begin{equation}
 0=
E^0_{JL}-E-
\label{final_simplified}
\sum\limits_{k \lambda  jl}
\frac{U_\lambda^2(k)
(K^{JL}_{jl\lambda})^2}
{\omega_k + E^0_{jl}-E},
\end{equation}
The last term in Eq.~\eqref{final_simplified} is the self-energy, from which one can obtain the Green's 
function and the spectral function.

\section{Normalization of the phonon populations}
\label{Appbeta}
In Sec.~\ref{sec:manipulating} we presented the plots of phonon populations. The quantity $\tilde{\beta}_{\lambda 
\tilde{j}l}(k)$ defined in Eq.~(\ref{eq:beta}) simply follows from Eq.~(\ref{der2}) if one temporarily assumes for the calculation that the quasiparticle weight $Z_{\tilde J L M_J}=1$. This results in normalization condition of Eq.~(\ref{eq:normalization}) for $\beta_{\lambda 
\tilde{j}l}(k)$ and complementary normalization condition for $Z_{\tilde J L M_J}$:
\begin{equation}
Z_{\tilde J L M_J}=\frac{1}{1+\sum\limits_{k\lambda \tilde{j}l}|\tilde{\beta}_{\lambda 
\tilde{j}l}(k)|^2}.
\end{equation}
As we can see, the normalizations of both phonon population $\beta_{\lambda \tilde j l}(k)$ and quasiparticle weight $Z_{\tilde J L M_J}$ inevitably involve an integral of type $\int_0^\infty dk \frac{f(k)}{(k-k_0)^2}$ At the angulon instabilities, there is a pole and the integral is divergent. This physically results in $|\beta_{\lambda \tilde j l}(k)|^2$ close to one for a given combination of $\lambda, \tilde j, l$ parameters and the rest of $\beta_{\lambda \tilde j l}(k)$ coefficients and the quasiparticle weight $Z_{\tilde JLM_J}$ being close to 0. Although there exist techniques such as Hadamard regularization, in our problem we need to calculate the value of the integral, not its finite part.

Let us also note that the problem presented above does not influence the calculations of spectral functions. There, the integrals are of type $\int_0^\infty dk \frac{f(k)}{k-k_0}$. There still might be a pole, but as we calculate the spectral function, the form of Eq.~(\ref{SpecFunc}) causes us to introduce a small but finite imaginary part to the denominator and take the real part of the integral. Mathematically, we know from Sokhotski-Plemelj theorem:
\begin{equation}
\lim_{\epsilon\rightarrow 0^+}\int\limits_a^b dk \frac{f(k)}{k-k_0\pm i\epsilon}=
\mp i\pi f(k_0)+\mathcal{P}\int\limits_a^b dk \frac{f(k)}{k-k_0},
\end{equation}
where $\mathcal{P}$ denotes the Cauchy principal value and $a<k_0<b$, that by taking $\epsilon$ small enough, in our numerical method we approach the Cauchy principal value of the integral. Hence, had we been able to calculate the analytical form of spectral function from Eq.~(\ref{SpecFunc}), it would differ from the one obtained numerically in this paper (with an introduction of a finite imaginary part) only by sharper spectral features.  

\twocolumngrid

\end{document}